\RequirePackage[l2tabu, orthodox]{nag}
\RequirePackage{snapshot}

\documentclass[10pt,onecolumn]{extarticle}

\makeatletter
\if@twocolumn
  \usepackage[dvips,letterpaper,top=0.5in, bottom=0.5in, left=0.75in, right=0.5in,includefoot,heightrounded]{geometry}
\else
  \usepackage[dvips,letterpaper,margin=1in,includefoot,heightrounded]{geometry}
\fi

\usepackage{srcltx}

\usepackage{amsmath}
\usepackage{amssymb,amsfonts}

\usepackage{abstract}

\usepackage{graphicx}
\usepackage{epsfig}
\usepackage[usenames,dvipsnames]{color}
\usepackage[usenames,dvipsnames]{xcolor}
\usepackage{subfigure}

\usepackage{booktabs}

\usepackage{setspace}
\usepackage{flushend}
\usepackage{multicol}

\usepackage{cite}
\usepackage{url}
\usepackage[normalem]{ulem}

\usepackage{enumerate}

\usepackage{hyperref}

\usepackage{lipsum}

\usepackage[sc,tiny,center]{titlesec}

\usepackage{microtype}

\makeatletter

\newcommand{\printtitle}{%
\makeatletter
\if@twocolumn

\twocolumn[%
  \maketitle
  \begin{onecolabstract}
    \myabstract
  \end{onecolabstract}
  \begin{center}
    \small
    \textbf{Keywords}
    \\\medskip
    \mykeywords
  \end{center}
  \bigskip
]
\saythanks
\else
  \maketitle
  \begin{onecolabstract}
    \myabstract
  \end{onecolabstract}
  \begin{center}
    \small
    \textbf{Keywords}
    \\\medskip
    \mykeywords
  \end{center}
  \bigskip
  \onehalfspacing
\fi
\makeatother
}

\title{%
DCT-like Transform for Image Compression
\\
Requires 14 Additions Only
}

\author{F.~M.~Bayer%
\thanks{%
F.~M.~Bayer
is with
the
Departamento de Estat\'istica,
Universidade Federal de Santa Maria,
RS, Brazil.
E-mail:
\protect\url{bayer@ufsm.br}
}
\and
R.~J.~Cintra%
\thanks{%
R.~J.~Cintra is with
the Signal Processing Group,
Departamento de Estat\'{\i}stica,
Universidade Federal de Pernambuco.
E-mail:
\protect\url{rjdsc@stat.ufpe.org}
}
}

\date{}

\newcommand{\myabstract}{%
A low-complexity 8-point orthogonal approximate DCT is introduced.
The proposed transform requires no multiplications or bit-shift operations.
The derived fast algorithm requires only 14 additions,
less than any existing DCT approximation.
Moreover,
in several image compression scenarios,
the proposed transform could outperform the well-known signed DCT,
as well as state-of-the-art algorithms.
}

\newcommand{\mykeywords}{%
DCT Approximation,
Fast algorithms,
Image compression
}

\begin{document}

\printtitle

\section{Introduction}

The discrete cosine transform (DCT) is an essential tool in digital signal processing (DSP).
In recent years, signal processing literature has been populated
with low-complexity methods for the efficient computation
of the 8-point DCT~\cite{Lecuire2012}.
Prominent approximation-based techniques include
the
signed DCT (SDCT)~\cite{Haweel2001},
the
level 1 approximation by Lengwehasatit-Ortega~\cite{Leng2004},
the
Bouguezel-Ahmad-Swamy (BAS) series of algorithms~\cite{bas2008,bas2009,bas2011},
and the
DCT round-off approximation~\cite{cb2011}.

In general, the transformation matrix entries required by
approximate DCT methods are only $\{ 0, \pm 1/2, \pm1, \pm2\}$.
This implies null multiplicative complexity,
because the involved operations
can be implemented
exclusively by means
additions and bit-shift operations.

In this letter, we introduce a low-complexity
DCT approximation that required only 14 additions.
The proposed algorithm attains the lowest computational complexity
among available methods found in literature.
At the same time, the proposed transform
could outperform state-of-the-art approximations.

\section{Proposed transform}

The proposed approximation is based on the
approximate DCT introduced in~\cite{cb2011};
hereafter referred to as CB-2011 matrix.
After judiciously replacing elements of the
CB-2011 matrix with zeros,
we obtained the following matrix:
$$
\mathbf{T} =
\begin{bmatrix}
1 &\phantom{-}1 &\phantom{-}1 & \phantom{-}1 & \phantom{-}1 &\phantom{-}1 &\phantom{-}1 &\phantom{-}1 \\
1 &\phantom{-}0 &\phantom{-}0 & \phantom{-}0 & \phantom{-}0 &\phantom{-}0 &\phantom{-}0 &-1 \\
1 &\phantom{-}0 &\phantom{-}0 &-1            & -1           &\phantom{-}0 &\phantom{-}0 &\phantom{-}1 \\
0 &\phantom{-}0 &-1           &\phantom{-}0  & \phantom{-}0 &\phantom{-}1 &\phantom{-}0 &\phantom{-}0 \\
1 &-1           &-1           & \phantom{-}1 & \phantom{-}1 &-1           &-1           &\phantom{-}1 \\
0 &-1           &\phantom{-}0 & \phantom{-}0 & \phantom{-}0 &\phantom{-}0 &\phantom{-}1 &\phantom{-}0 \\
0 &-1           &\phantom{-}1 & \phantom{-}0 & \phantom{-}0 &\phantom{-}1 &-1           &\phantom{-}0 \\
0 &\phantom{-}0 &\phantom{-}0 &-1            & \phantom{-}1 &\phantom{-}0 &\phantom{-}0 &\phantom{-}0
\end{bmatrix}
.
$$
Above matrix furnishes the approximate DCT expressed by:
$\hat{\mathbf{C}}= \mathbf{D} \cdot \mathbf{T}$,
where
$
\mathbf{D} = \mathrm{diag}
\left(
\frac{1}{\sqrt{8}},
\frac{1}{\sqrt{2}},
\frac{1}{2},
\frac{1}{\sqrt{2}},
\frac{1}{\sqrt{8}},
\frac{1}{\sqrt{2}},
\frac{1}{2},
\frac{1}{\sqrt{2}}
\right)
$.

The entries of $\mathbf{T}$ are $\{0, \pm 1\}$.
This is an attestation of its null multiplicative complexity. Moreover, bit-shift operations are fully absent.
Not only $\hat{\mathbf{C}}$ inherits the low computational complexity of $\mathbf{T}$,
but it is also orthogonal.
In terms of complexity assessment,
matrix~$\mathbf{D}$ may not introduce any computational overhead~\cite{Leng2004,bas2008,bas2009,bas2011,cb2011}.
In image compression, the DCT operation is a pre-processing step
for subsequent coefficient quantisation.
In this context, matrix~$\mathbf{D}$, in the form of $\mathbf{D}^2$,
can be merged into the quantisation matrix.
Moreover, all elements of $\mathbf{D}^2$ are negative powers of two $\{1/2,1/4,1/8 \}$. Therefore,  any implementation of the quantisation step for  the exact DCT can be easily adapted to the proposed method by adequately bit-shifting the elements of the quantisation matrix.

\begin{figure}
\centering
\input{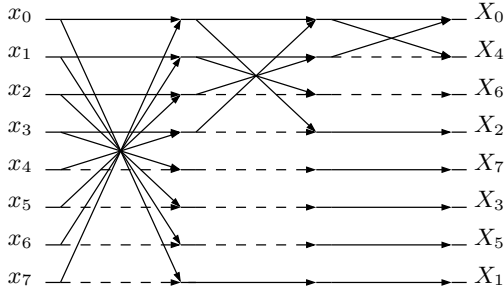}
\caption{
Signal flow graph for $\mathbf{T}$.
Input data $x_n$, $n=0,1,\ldots,7$,
relates to output $X_k$, $k=0,1,\ldots,7$,
according to $\mathbf{X} = \mathbf{T} \cdot \mathbf{x}$.
Dashed arrows represent multiplication by $-1$.}
\label{diagram}
\end{figure}

A fast algorithm based on sparse matrix factorization leads to $\mathbf{T}=\mathbf{P}\cdot \mathbf{A}_3 \cdot \mathbf{A}_2 \cdot \mathbf{A_1}$,
where:
\begin{align*}
&\mathbf{A}_1 =
\begin{bmatrix}
\begin{smallmatrix}
1 &\phantom{-}0 &\phantom{-}0 & \phantom{-}0 & \phantom{-}0 &\phantom{-}0 &\phantom{-}0 &\phantom{-}1 \\
0 &\phantom{-}1 &\phantom{-}0 & \phantom{-}0 & \phantom{-}0 &\phantom{-}0 &\phantom{-}1 &\phantom{-}0 \\
0 &\phantom{-}0 &\phantom{-}1 & \phantom{-}0 & \phantom{-}0 &\phantom{-}1 &\phantom{-}0 &\phantom{-}0 \\
0 &\phantom{-}0 &\phantom{-}0 & \phantom{-}1 & \phantom{-}1 &\phantom{-}0 &\phantom{-}0 &\phantom{-}0 \\
0 &\phantom{-}0 &\phantom{-}0 & \phantom{-}1 & -1			&\phantom{-}0 &\phantom{-}0 &\phantom{-}0 \\
0 &\phantom{-}0 &\phantom{-}1 & \phantom{-}0 & \phantom{-}0 & -1          &\phantom{-}0 &\phantom{-}0 \\
0 &\phantom{-}1 &\phantom{-}0 & \phantom{-}0 & \phantom{-}0 &\phantom{-}0 & -1          &\phantom{-}0 \\
1 &\phantom{-}0 &\phantom{-}0 & \phantom{-}0 & \phantom{-}0 &\phantom{-}0 &\phantom{-}0 & -1
\end{smallmatrix}
\end{bmatrix}\!, \;
{%
\mathbf{A}_2 =
\begin{bmatrix}
\begin{smallmatrix}
1 &\phantom{-}0 &\phantom{-}0 & \phantom{-}1 & \phantom{-}0 &\phantom{-}0 &\phantom{-}0 &\phantom{-}0 \\
0 &\phantom{-}1 &\phantom{-}1 & \phantom{-}0 & \phantom{-}0 &\phantom{-}0 &\phantom{-}0 &\phantom{-}0 \\
0 &\phantom{-}1 & -1          & \phantom{-}0 & \phantom{-}0 &\phantom{-}0 &\phantom{-}0 &\phantom{-}0 \\
1 &\phantom{-}0 &\phantom{-}0 & -1           & \phantom{-}0 &\phantom{-}0 &\phantom{-}0 &\phantom{-}0 \\
0 &\phantom{-}0 &\phantom{-}0 & \phantom{-}0 & -1           &\phantom{-}0 &\phantom{-}0 &\phantom{-}0 \\
0 &\phantom{-}0 &\phantom{-}0 & \phantom{-}0 & \phantom{-}0 & -1          &\phantom{-}0 &\phantom{-}0 \\
0 &\phantom{-}0 &\phantom{-}0 & \phantom{-}0 & \phantom{-}0 &\phantom{-}0 & -1          &\phantom{-}0 \\
0 &\phantom{-}0 &\phantom{-}0 & \phantom{-}0 & \phantom{-}0 &\phantom{-}0 &\phantom{-}0 &\phantom{-}1
\end{smallmatrix}
\end{bmatrix},}\\
&\mathbf{A}_3 =
\begin{bmatrix}
\begin{smallmatrix}
1 &\phantom{-}1 &\phantom{-}0 & \phantom{-}0 & \phantom{-}0 &\phantom{-}0 &\phantom{-}0 &\phantom{-}0 \\
1 & -1          &\phantom{-}0 & \phantom{-}0 & \phantom{-}0 &\phantom{-}0 &\phantom{-}0 &\phantom{-}0 \\
0 &\phantom{-}0 &-1           & \phantom{-}0 & \phantom{-}0 &\phantom{-}0 &\phantom{-}0 &\phantom{-}0 \\
0 &\phantom{-}0 &\phantom{-}0 & \phantom{-}1 & \phantom{-}0 &\phantom{-}0 &\phantom{-}0 &\phantom{-}0 \\
0 &\phantom{-}0 &\phantom{-}0 & \phantom{-}0 & \phantom{-}1 &\phantom{-}0 &\phantom{-}0 &\phantom{-}0 \\
0 &\phantom{-}0 &\phantom{-}0 & \phantom{-}0 & \phantom{-}0 &\phantom{-}1 &\phantom{-}0 &\phantom{-}0 \\
0 &\phantom{-}0 &\phantom{-}0 & \phantom{-}0 & \phantom{-}0 &\phantom{-}0 &\phantom{-}1 &\phantom{-}0 \\
0 &\phantom{-}0 &\phantom{-}0 & \phantom{-}0 & \phantom{-}0 &\phantom{-}0 &\phantom{-}0 &\phantom{-}1
\end{smallmatrix}
\end{bmatrix}\!,\;
\mathbf{P} =
\begin{bmatrix}
\begin{smallmatrix}
1 &\phantom{-}0 &\phantom{-}0 & \phantom{-}0 & \phantom{-}0 &\phantom{-}0 &\phantom{-}0 &\phantom{-}0 \\
0 &\phantom{-}0 &\phantom{-}0 & \phantom{-}0 & \phantom{-}0 &\phantom{-}0 &\phantom{-}0 &\phantom{-}1 \\
0 &\phantom{-}0 &\phantom{-}0 & \phantom{-}1 & \phantom{-}0 &\phantom{-}0 &\phantom{-}0 &\phantom{-}0 \\
0 &\phantom{-}0 &\phantom{-}0 & \phantom{-}0 & \phantom{-}0 &\phantom{-}1 &\phantom{-}0 &\phantom{-}0 \\
0 &\phantom{-}1 &\phantom{-}0 & \phantom{-}0 & \phantom{-}0 &\phantom{-}0 &\phantom{-}0 &\phantom{-}0 \\
0 &\phantom{-}0 &\phantom{-}0 & \phantom{-}0 & \phantom{-}0 &\phantom{-}0 &\phantom{-}1 &\phantom{-}0 \\
0 &\phantom{-}0 &\phantom{-}1 & \phantom{-}0 & \phantom{-}0 &\phantom{-}0 &\phantom{-}0 &\phantom{-}0 \\
0 &\phantom{-}0 &\phantom{-}0 & \phantom{-}0 & \phantom{-}1 &\phantom{-}0 &\phantom{-}0 &\phantom{-}0
\end{smallmatrix}
\end{bmatrix}.
\end{align*}
Signal flow graph for $\mathbf{T}$ is shown in Fig.~\ref{diagram}.

Arithmetic complexity assessment and comparisons with
state-of-the-art DCT approximations
are shown in Table~\ref{complex}.
Demanding only 14 additions, proposed transform ${\hat{\mathbf{C}}}$ possesses
$22.2\%$, $30.0\%$, and $41.7\%$ lower arithmetic costs than
the BAS-2009 transform~\cite{bas2009},
the BAS-2011 transform~\cite{bas2011},
and
the SDCT, respectively.
Notice that the BAS-2011 transform is the most recent algorithm in the BAS series.

DSP literature contains the DCT approximation
described in~\cite{bb2011}, which is
claimed to require 16 additions.
However, we could not reproduce the performance results
shown in~\cite{bb2011}. Indeed, contrary to~\cite{bb2011},
such approximation could not be verified to be orthogonal.
Thus, we could not consider~\cite{bb2011} for any meaningful comparison.

\begin{table}
\centering
\caption{Arithmetic complexity analysis}
\label{complex}
\begin{tabular}{|l|c|c|c|c|}
\hline
Method & Add. & Mult. & Shifts & Total\\
\hline
Proposed transform
& 14 & 0 & 0 & 14 \\
\hline
SDCT~\cite{Haweel2001}
& 24 & 0 & 0 & 24 \\
\hline
Level 1 approximation~\cite{Leng2004}
& 24 & 0 & 2 & 26 \\
\hline
BAS-2008 transform~\cite{bas2008}
& 18 & 0 & 2 & 20 \\
\hline
BAS-2009 transform~\cite{bas2009}
& 18 & 0 & 0 & 18 \\
\hline
BAS-2011 transform~\cite{bas2011}
& 18 & 0 & 2 & 20 \\
\hline
CB-2011 transform~\cite{cb2011}
& 22 & 0 & 0 & 22\\
\hline
\end{tabular}
\end{table}

\section{Image compression}

To assess the performance of the proposed transform for image compression,
we used the methodology described in~\cite{Haweel2001}
and supported by~\cite{Leng2004,bas2008,bas2009,bas2011,cb2011}.
A set of 45 $512\times 512$ 8-bit greyscale images obtained
from a standard public image bank~\cite{usc} was considered.
We implemented the JPEG image compression technique for the 8$\times$8 matrix case.
Each image was divided into 8$\times$8 sub-blocks,
which were submitted to the two-dimensional transforms.
This computation furnished 64 coefficients in the approximate transform domain
for each sub-block.
According to the standard zigzag sequence %
only the $r$ initial coefficients in each block
were employed to reconstruct the image~\cite{cb2011}.
All the remaining coefficients were set to zero.
We adopted $2 \leq r \leq 45$,
which corresponds to compression ratios
between $96.875\%$ and $29.690\%$, respectively.
The inverse procedure was then applied to reconstruct the processed data
and image degradation was assessed.

For the sake of image compression performance assessment,
the peak signal-to-noise ratio (PSNR)
and mean square error (MSE) were utilized as figures of merit.
However, in contrast with the numerical experiments described in~\cite{Haweel2001,Leng2004,bas2008,bas2009,bas2011},
we adopted the average quality measure from all considered images
instead of the results obtained from particular images.
Thus, our analysis is more robust;
being less prone to variance effects and fortuitous input data.
Among available algorithms, we separate the SDCT~\cite{Haweel2001}
and BAS-2011~\cite{cb2011} for comparison.
The SDCT is a classic reference in the field~\cite{Brita2007}
and the BAS-2011 transform is the
most recent method in the BAS series of algorithms.
The parametric transform BAS-2011 was considered with
parameter $a=0.5$~\cite{bas2011}.

\begin{figure}
\centering
\subfigure[Average PSNR]
{\includegraphics[width=0.475\linewidth]{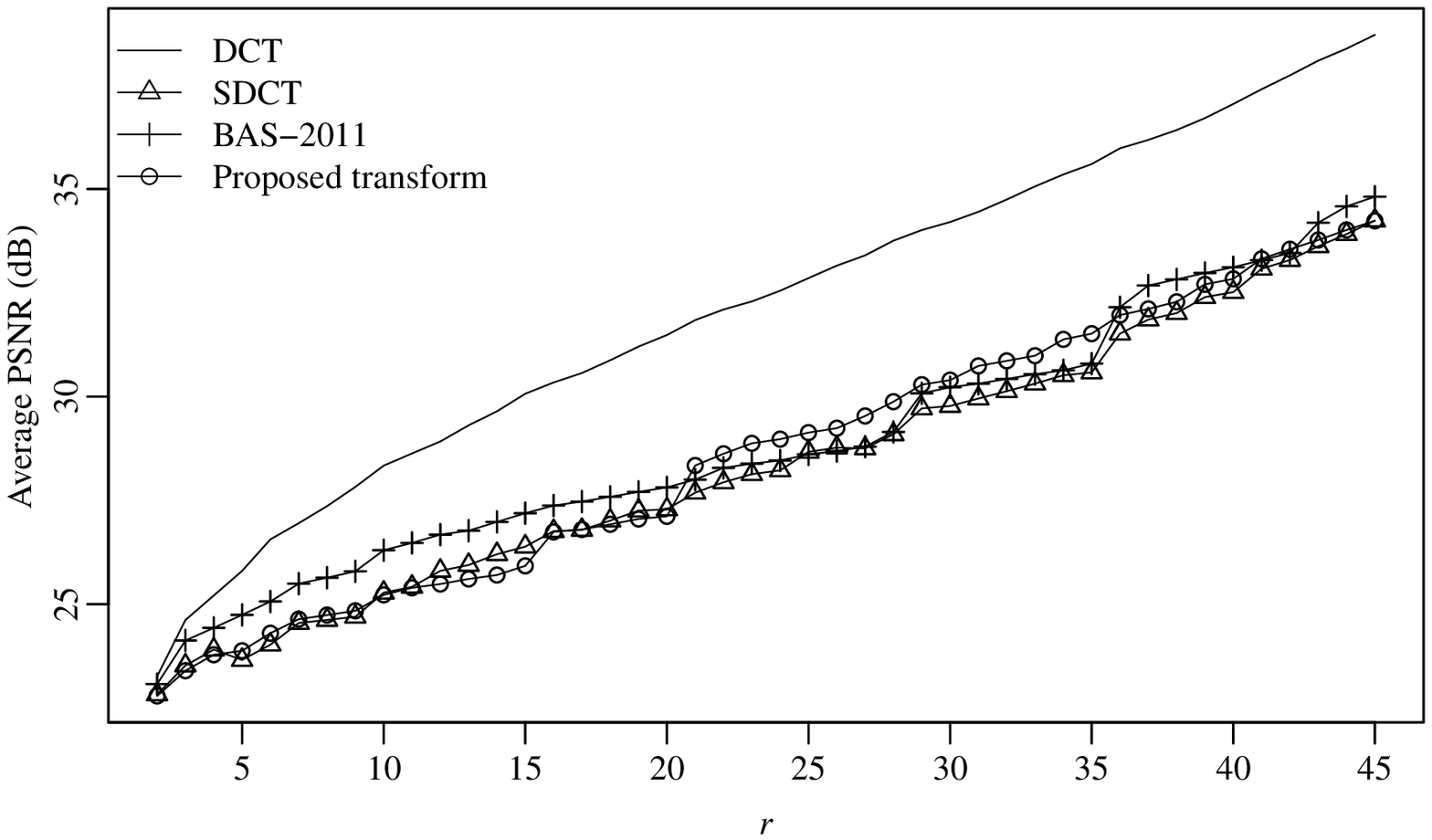} \label{psnr}}
\subfigure[Average MSE absolute percentage error relative to the DCT]
{\includegraphics[width=0.475\linewidth]{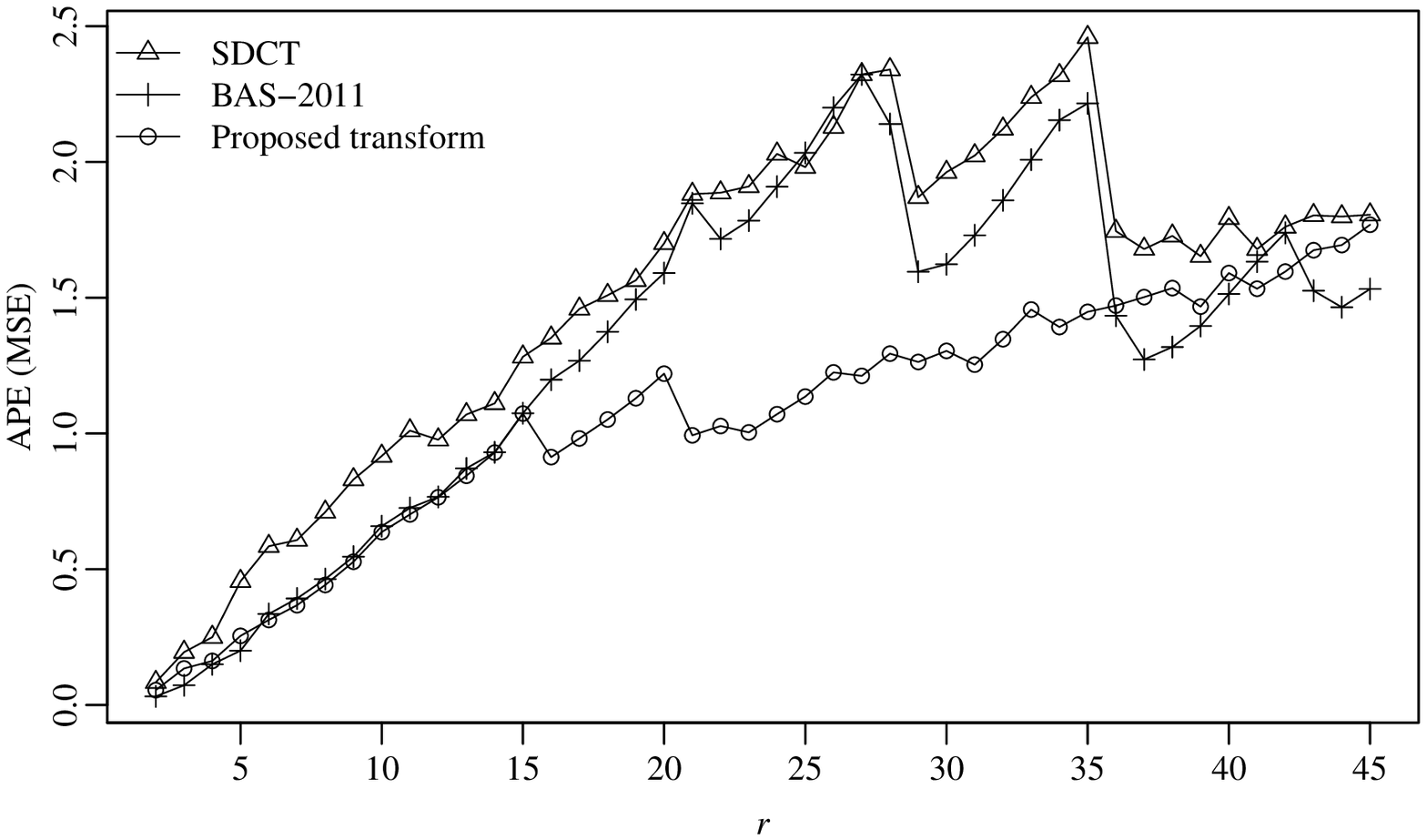} \label{ape_mse}}

\caption{Quality measures for several compression ratios.}
\end{figure}

Fig.~\ref{psnr} shows the resulting PSNR measures.
The proposed approximation~${\hat{\mathbf{C}}}$ has comparable
performance to the SDCT at high compression rates and
could indeed outperform it at low compression rates.
At the mid-range compression ratios ($20<r \leq 35$),
the proposed matrix outperformed both the BAS-2011 transform and the SDCT.
This similar result could be achieved in despite of requiring
only $70.0\%$ and $58.3\%$ of the arithmetic cost of
the BAS-2011 transform and the SDCT, respectively.
Fig.~\ref{ape_mse} depicts the absolute percentage error (APE)
relative to the exact DCT for the average MSE.
According to this metric,
the proposed approximation
led to a better performance at
compression ratios ranging from $90.625\%$ ($r=6$)
to
$45.310\%$ ($r=35$),
in which popular compression ratios are included.

\begin{figure}
\centering
\subfigure[DCT (PSNR=37.21)]
{\includegraphics[width=0.3\linewidth]{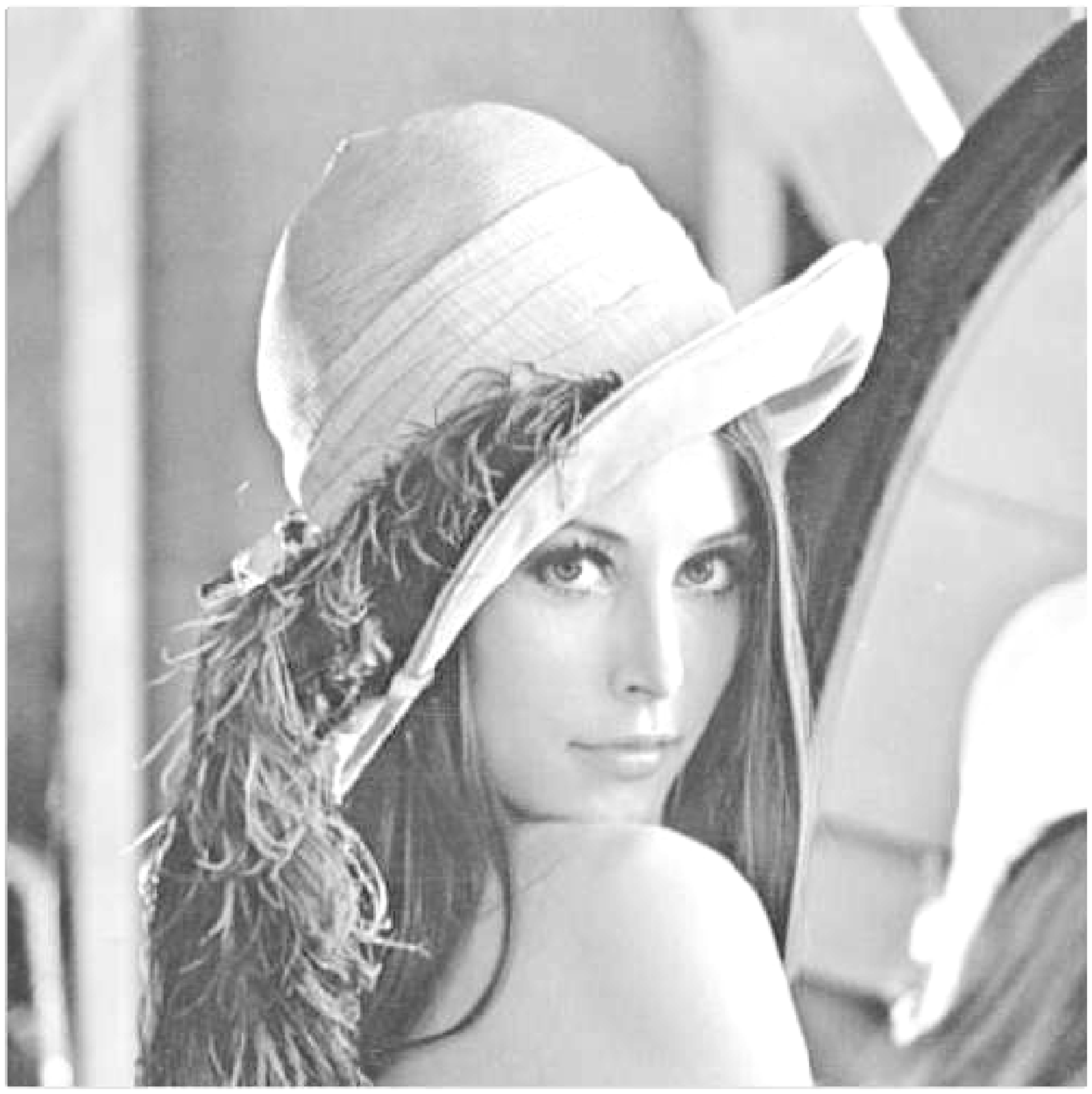}}
\subfigure[Proposed (PSNR=31.44)]
{\includegraphics[width=0.3\linewidth]{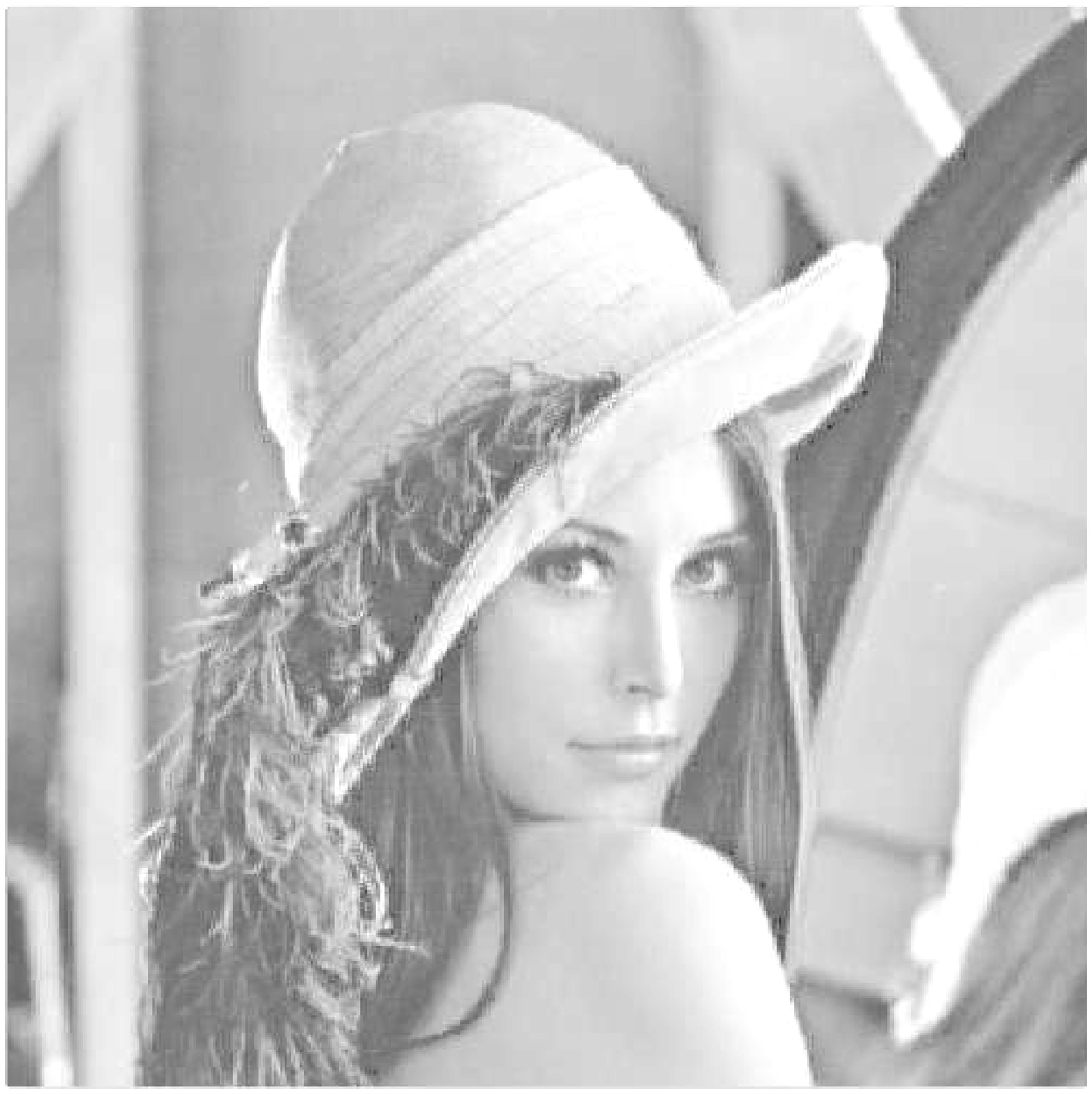}}

\subfigure[BAS-2011 (PSNR=31.33)]
{\includegraphics[width=0.3\linewidth]{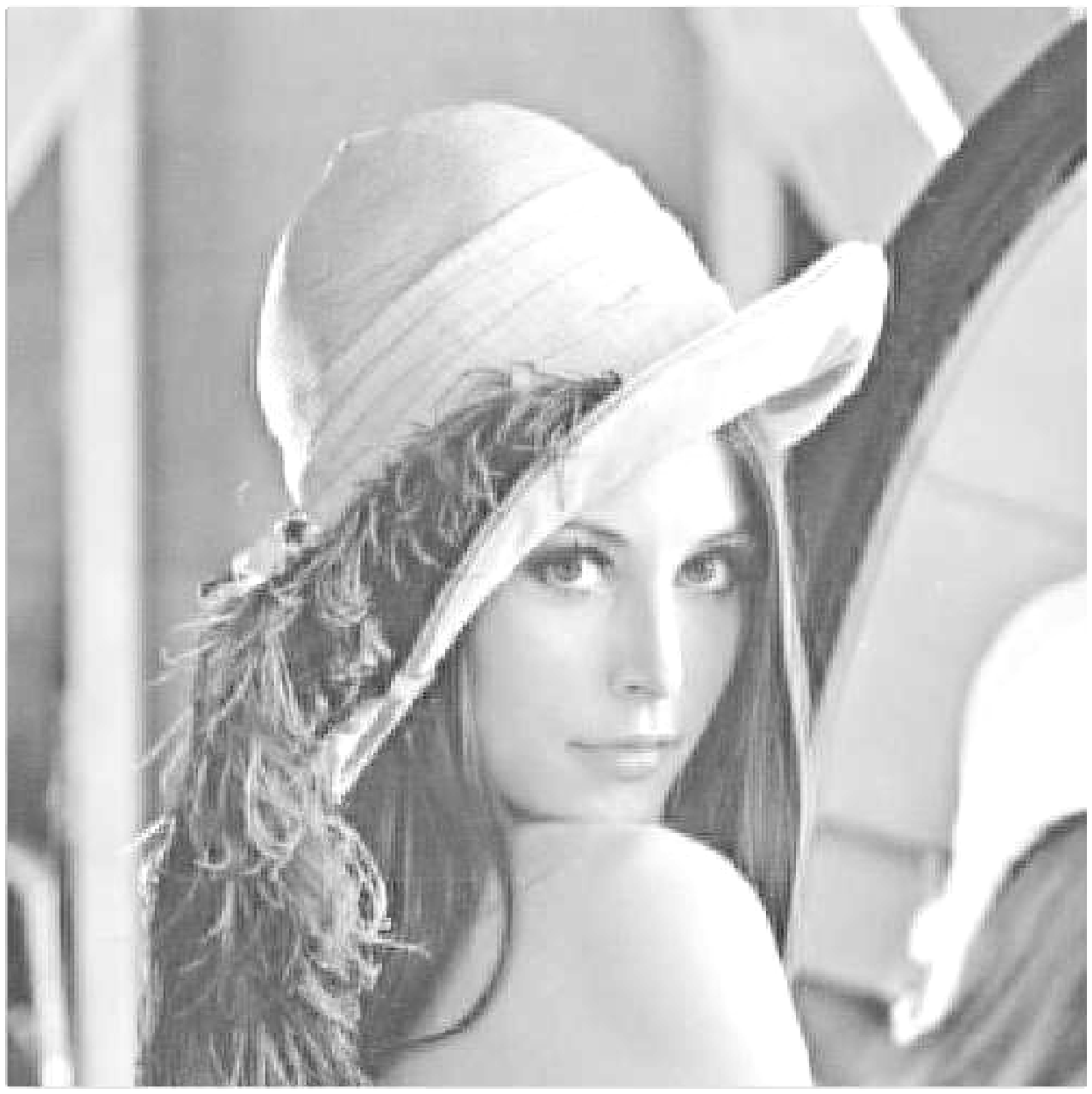}}
\subfigure[SDCT (PSNR=31.25)]
{\includegraphics[width=0.3\linewidth]{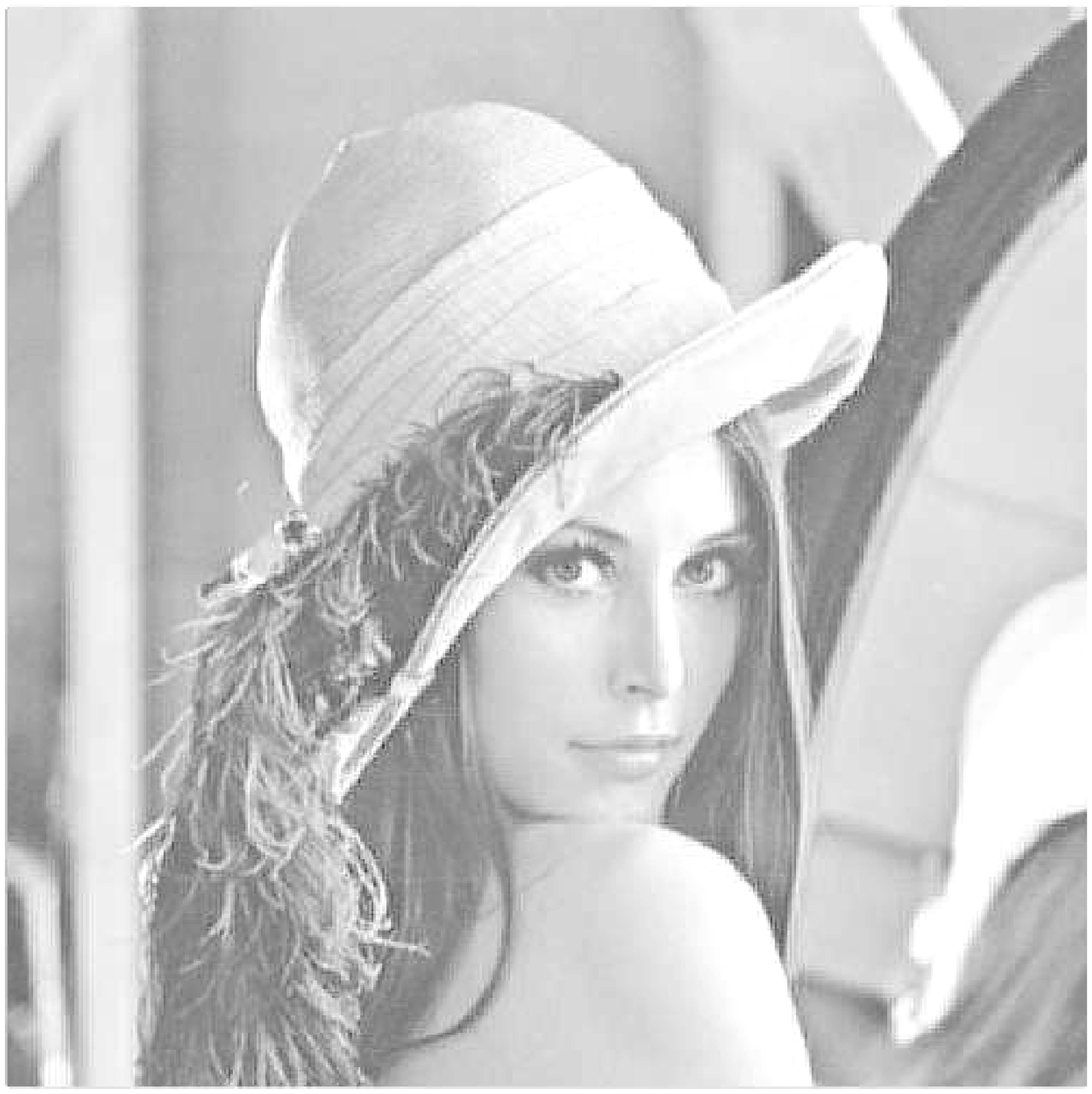}}

\caption{Compressed Lena image using (a) the DCT, (b) the proposed transform, (c) the BAS-2011 transform, and (d) the SDCT, for $r=25$.}
\label{lena}
\end{figure}

Fig.~\ref{lena} shows a qualitative comparison
including
the DCT,
the proposed transform,
the BAS-2011 transform,
and the SDCT.
A $60.937\%$ compression ($r=25$) was applied to the standard Lena image.
Proposed transform offered results that are comparable
to those furnished by the exact DCT.

\section{Conclusion}

This letter introduced an 8-point
transform suitable for image compression.
The proposed transform requires only 14 additions
and has comparable or better image compression performance
than the classic SDCT and the state-of-the-art BAS-2011 transform.

\end{document}